\documentclass[journal,comsoc]{IEEEtran}

\usepackage[T1]{fontenc}% optional T1 font encoding
\usepackage{cite,amssymb,color,textcomp,tabularx}
\usepackage{graphicx,epstopdf}
\usepackage{amsmath}
\usepackage{bm}
\usepackage{algorithm,algorithmic}
\usepackage[caption=false,font=footnotesize]{subfig}
\usepackage{stfloats}
\usepackage{multirow}%tabular cells with multiple rows
\usepackage{enumerate,extarrows}
\usepackage{dsfont}% 1(.) indicator function
\usepackage[mathcal]{euscript}

\newtheorem{theorem}{\underline{Theorem}}%[section]
\newtheorem{lemma}{\underline{Lemma}}%[section]
\newtheorem{corollary}{\underline{{Corollary}}}%[section]
\newtheorem{proposition}{\underline{{Proposition}}}%[section]

\def \ln {\mathrm{ln}}
\def \Ei {\mathrm{Ei}}

\def \st {\mathrm{s.t.}}
\def \P {\mathbb{P}}
\def \E {\mathbb{E}}
\def \C {\mathbb{C}}

\def \maxp {\mathop{\mathrm{maximize}}}

\newcommand{\mb}[1]{\mathbf{#1}}

\begin{document}
\title{Secure Short-Packet Communications for Mission-Critical IoT Applications}
\author{
	Hui-Ming~Wang,~%
	Qian~Yang,~%
	Zhiguo~Ding,~%
	and~H.~Vincent~Poor%
\thanks{H.-M. Wang and Q. Yang are with the School of Electronic and Information Engineering, Xi'an Jiaotong University, Xi'an 710049, China, and also with the Ministry of Education Key Laboratory for Intelligent Networks and Network Security, Xi'an Jiaotong University, Xi'an 710049, China (e-mail: xjbswhm@gmail.com; yangq36@gmail.com).}%
\thanks{Z. Ding and H. V. Poor are with the Department of Electrical Engineering,  Princeton University, Princeton, NJ 08544, USA. Z. Ding is  also with the School of Electrical and Electronic Engineering, the University of Manchester, Manchester M13 9PL, U.K. (e-mail: zhiguo.ding@manchester.ac.uk; poor@princeton.edu).}%
	}
\maketitle

\begin{abstract}
	In pervasive Internet of Things (IoT) applications, the use of short packets is expected to meet the stringent latency requirement in ultra-reliable low-latency communications; however, the incurred security issues and the impact of finite blocklength coding on the physical-layer security have not been well understood.
	This paper comprehensively investigates the performance of secure short-packet communications in a mission-critical IoT system with an external multi-antenna eavesdropper.
	An analytical framework is proposed to approximate the average achievable secrecy throughput of the system with finite blocklength coding.
	To gain more insight, a simple case with a single-antenna access point (AP) is considered first, in which the secrecy throughput is approximated in a closed form.
	Based on that result, the optimal blocklengths to maximize the secrecy throughput with and without the reliability and latency constraints, respectively, are derived.
	For the case with a multi-antenna AP, following the proposed analytical framework, closed-form approximations for the secrecy throughput are obtained under both beamforming and artificial-noise-aided transmission schemes.
	Numerical results verify the accuracy of the proposed approximations and illustrate the impact of the system parameters on the tradeoff between transmission latency and reliability under the secrecy constraint.
\end{abstract}

\begin{IEEEkeywords}
	Finite blocklength, physical-layer security, ultra-reliable low-latency communications, short-packet communications.
\end{IEEEkeywords}

%\newpage
%\linespread{1.5}

%\IEEEPARstart{T}{he past}
\section{Introduction}\label{sec_intro}
%background of mtc, urllc
The past three decades have witnessed the prominent technical advancement of conventional mobile broadband (MBB) characterized by high throughput and capacity.
In the upcoming fifth-generation (5G) era, the emphasis on the support of Internet of Things (IoT) functionality becomes pronounced, since the IoT is expected to enable new applications and intelligent decision making by bridging diverse technologies and connecting physical objects together\cite{Durisi2016,Schulz2017,Al-Fuqaha2015}.
%Ranging from tiny sensors to self-driving vehicles, the ubiquitous IoT applications in 5G systems are characterized by stringent constraints on latency and reliability, huge number of connections, high scalability, and high energy-efficiency.
To support the IoT in 5G, two novel service categories, namely, ultra-reliable low-latency communications (uRLLC) and massive machine-type communications (mMTC), have been introduced\cite{Popovski2014}.
For the use case of mMTC, 5G solutions aim to provide wireless connectivity to tens of billions of usually low-cost energy-constrained machine-type devices such as sensors and wearable devices\cite{Bockelmann2016}.
In contrast, uRLLC requirements are closely connected with mission-critical machine-type applications with the emphasis on the stringent constraints of reliability and latency, as typically required by intelligent transportation and industry automation\cite{Schulz2017}.

%intro pls
With the advent of IoT applications such as human-centric intelligent wearables and smart healthcare, unprecedented security challenges, such as privacy leakage, financial loss, and malicious invasion, arise\cite{Liu2016e,He2018}.
Due to the broadcast nature of wireless communications, IoT communication systems are particularly vulnerable to eavesdropping\cite{Barki2016}.
Conventionally, security for IoT is strengthened through efficient cryptography at the upper layers of the communication protocol stack\cite{Granjal2015}.
However, secret key generation, distribution, and management can be complicated and difficult to implement for large-scale peer-to-peer IoT networks with many low-complexity nodes like sensors\cite{Poor2012}.
Compared with cryptographic technologies used at the upper layers, physical-layer security, as an alternative to cryptography, is more appealing since perfect secrecy is accomplished by exploiting the random nature of wireless channels\cite{Poor2017}.
%The information-theoretic study of physical-layer security is based on the pioneering work of Wyner in \cite{Wyner1975}, where the wiretap channel model was originally introduced, and on subsequent generalization to general discrete memoryless channels \cite{Csiszar1978} and Gaussian wiretap channels \cite{Yan-Cheong1978}.
A fundamental quantity of interest in physical-layer security is the secrecy capacity.
%, which refers to the maximal rate at which information can be both securely and reliably transmitted in the presence of eavesdroppers.
%the maximal secrecy transmission rate is called the secrecy capacity, which characterizes the rate of the information that can be both securely and reliably transmitted against eavesdroppers.
Provided that the transmission rate is less than the secrecy capacity and the coding blocklength is sufficiently large, both the decoding error probability at the legitimate receiver and the amount of information leaked to the eavesdropper can be made arbitrarily small\cite{Bloch2011,Poor2017}.

%intro short packet
Despite the aforementioned security challenges, applying physical-layer security measures in mission-critical IoT applications is nontrivial.
As pointed out in \cite{Durisi2016}, uRLLC and mMTC scenarios feature short-packet communications, and supporting short-packet transmission is the primary challenge in these applications.
From \cite{Schulz2017}, \cite{Bockelmann2016} and \cite{Boccardi2014}, it is commonly known that short packets used in IoT applications can potentially go down to hundreds of bits.
In this regard, channel codes or wiretap codes conventionally adopted in cellular networks cannot be used for short-packet communications. This is because these codes are constructed for long packets (e.g., packet size $\gg 10^3$ bytes) so as to approach Shannon's channel capacity or secrecy capacity.
Furthermore, the classical information-theoretic results for performance analysis are no longer valid for short packets, due to the fact that the law of large number cannot be leveraged for the case with finite packet length (coding blocklength)\cite{Durisi2016}.
This calls for a fundamentally different design for secure short-packet communications in 5G IoT applications compared with conventional data-rate-centric systems.

\subsection{Related Work and Motivation}\label{subsec_related}
From the perspective of the physical layer, the design of uRLLC services is arguably more challenging than mMTC due to the fact that the two stringent constraints in mission-critical IoT applications are conflicting.
On one hand, to achieve low latency the use of short packets becomes mandated, but the channel coding gain can be severely degraded, which means that it becomes difficult to ensure reliability.
On the other hand, more resources are required for re-transmission and redundancy to improve reliability, but they incur greater latency.
Therefore, how to carefully balance the two constraints and realize the goal of uRLLC with the aid of physical-layer security is of significant interest for secure mission-critical IoT applications.
%However, the research on this field is still in its infancy.

Due to its potential practical applications, physical-layer security has been extensively investigated in various communication systems \cite{Mukherjee2014,Mukherjee2015,Yang2015a,Liu2017d,Poor2017}.
However, infinite blocklength is assumed by most of the existing work on physical-layer security. This assumption no longer holds for IoT applications that require short-packet transmission. % to achieve a low latency.

During the past few years, prominent work has been done to investigate the performance of short-packet communications from an information-theoretic perspective\cite{Polyanskiy2010,Yang2014,Durisi2016a}.
In \cite{Polyanskiy2010}, the maximal achievable channel coding rate was investigated at a given blocklength and error probability. It is shown in \cite{Polyanskiy2010} that, unlike the classical Shannon formulation, the reliability cannot be arbitrarily high in the finite blocklength regime, and the incurred penalty on the maximal transmission rate can be high.
Under the case of multiple-input multiple-output (MIMO) fading channels with finite blocklength coding, the maximal achievable rate was analyzed in \cite{Yang2014}, and the tradeoff between reliability, throughput, and latency was investigated in \cite{Durisi2016a}.
These information-theoretic results were then used to study the impact of finite blocklength coding on incremental redundancy based hybrid automatic repeat request (HARQ)\cite{Makki2014}, wireless energy transfer\cite{Makki2016,Lopez2017}, two-way relaying\cite{Gu2018}, and non-orthogonal multiple access (NOMA)\cite{Yu2018,Sun2018} systems.
The tradeoff between the sum rate and error probability in a multi-user downlink system with finite blocklength coding was investigated in \cite{Haghifam2017}.
In addition, for the case with no channel state information (CSI) at the transmitter, the tradeoff between energy efficiency and spectral efficiency with finite blocklength coding was analyzed in \cite{Mary2016}.
However, the above work does not take the transmission secrecy into consideration.

%intro pls+short packet for urllc
So far, there is only limited work devoted to investigating the secrecy rate under the case of finite blocklength \cite{Hayashi2006,Tan2012,Yassaee2013,Yang2016d,Yang2017d}.
In \cite{Hayashi2006}, general achievability bounds for wiretap channels with finite blocklength coding were obtained.
Subsequently, much effort has been put to improving the bounds\cite{Tan2012,Yassaee2013}.
For given reliability and secrecy constraints under the finite blocklength case, the tightest bounds and the second-order coding rate for discrete memoryless and Gaussian wiretap channels were obtained in \cite{Yang2016d} and \cite{Yang2017d}.
However, to the best of the authors' knowledge no results are available on analyzing practical system performance with these information-theoretic results, nor has there been a comprehensive study of the secrecy system throughput with finite blocklength coding.
Furthermore, how to design the blocklength in order to balance the latency-reliability tradeoff under the secrecy constraint remains unclear.
The above issues motivate our work.

\subsection{Our Work and Contributions}\label{subsec_our}
%in this paper
In this paper, we comprehensively investigate the average achievable secrecy throughput of short-packet communications in a secure IoT system, in which an access point (AP) aims to securely communicate with an actuator in the presence of a multi-antenna eavesdropper.
%Using the result in \cite{Yang2016d,Yang2017d},
We propose an analytical framework to evaluate the secrecy throughput for both the single-antenna and multi-antenna AP cases with finite-blocklength transmission.
%The performance analysis is carried out under the scenarios where the antenna number of the AP is one and larger than one, respectively, and
Based on the derived approximations, the impact of blocklength on the latency-reliability tradeoff under the secrecy constraint is studied, and the secrecy throughput is further optimized.
Note that a comprehensive theoretical performance study of the secrecy system throughput with finite blocklength coding is provided in our paper, and the obtained results can be used to guide the practical parameter design in 5G new radio systems. Specifically, the novelties and main contributions of this paper are summarized as follows:

\begin{enumerate}[1)]
	\item The information-theoretic results on finite-blocklength bounds for wiretap channels are leveraged to analyze the average achievable secrecy throughput of a practical IoT system with a multi-antenna eavesdropper. Through the proposed analytical framework, closed-form approximations for the secrecy throughput are derived for both cases in which the AP has one and multiple antennas.
	\item For the case with a single-antenna AP, the impact of blocklength on the latency-reliability tradeoff under the secrecy constraint is analyzed. In particular, the optimal blocklength in terms of secrecy throughput maximization is derived, and the impacts of the system parameters and of the reliability and latency constraints on the optimum are further analyzed.
	\item For the case with a multi-antenna AP, the AP can employ beamforming or artificial-noise-aided (AN-aided) transmission schemes for reliability and security enhancements.
	%For the case with a multi-antenna AP, both the beamforming and artificial-noise-aided (AN-aided) transmission schemes are investigated for reliability and security enhancements.
	We obtain a closed-form expression for the secrecy throughput under this case, which holds for cases with arbitrary numbers of antennas at the AP and the eavesdropper.
	\item Numerical results are provide to verify the accuracy of the proposed approximations and uncover the impact of the system parameters on the tradeoff between transmission latency and reliability under the secrecy constraint. It is shown that adding more antennas at the AP can compensate for the performance loss incurred by short-packet communications in a way. % in the large-blocklength regime.
	%equipping multiple antennas at the transmitter with the AN-aided scheme can significantly improve the secrecy throughput with finite blocklength coding.
\end{enumerate}

\subsection{Organization and Notation}
The rest of this paper is organized as follows: In Sections \ref{sec_model}, we present the system model and the performance metrics for the secure short-packet transmission.
In Sections \ref{sec_single_performance} and \ref{sec_single_opt}, we investigate the secrecy throughput and its optimization for the case with a single-antenna AP, respectively, while the multi-antenna case is studied in Section \ref{sec_multi}.
Numerical simulations and analysis are presented in Section \ref{sec_sim} before the conclusions are drawn in Section \ref{sec_conclusion}.

\emph{Notation:}
$\mathbf{A}^T$ and $\mathbf{A}^H$ represent the transpose and conjugate transpose of a matrix $\mathbf{A}$, respectively.
$\mathbb{E}\{\cdot\}$ denotes expectation, and $\mathbf{I}_K$ denotes a $ K $-dimensional identity matrix.
$\mathbf{x}\sim \mathcal{CN}(\bm{\mu},\bm{\Sigma})$ means that $\mathbf{x}$ is a random vector following a complex circular Gaussian distribution with mean $\bm{\mu}$ and covariance $\bm{\Sigma}$.
$X\sim\mathrm{Exp}(\lambda)$ denotes an exponentially distributed random variable with rate $\lambda$, and $X\sim\mathrm{Gamma}(k,\theta)$ denotes a gamma-distributed random variable with shape $k$ and scale $\theta$.
$ \Ei(x) $ is the exponential integral function \cite[eq. (8.211.1)]{Gradshteyn2007}, $\Gamma(x)$ is the gamma function \cite[eq. (8.310.1)]{Gradshteyn2007}, $\gamma(k,x)$ is the lower incomplete gamma function \cite[eq. (8.350.1)]{Gradshteyn2007}, and $\Gamma(k,x)$ is the upper incomplete gamma function \cite[eq. (8.350.2)]{Gradshteyn2007}.
The ceiling and floor operations are denoted by $ \lceil\cdot\rceil $ and $ \lfloor\cdot\rfloor $, respectively.

\section{System Model and Performance Metric}\label{sec_model}

\subsection{System Model}\label{subsec_sys_model}
\begin{figure}[t]
	\centering
	\includegraphics[width=2.25in]{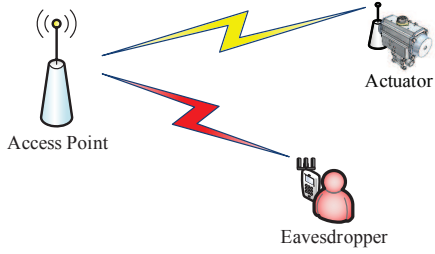}
	\caption{A secure IoT communication system in which an access point transmits confidential short-packet information to a trusted actuator in the presence of eavesdropping.
	}
	\label{fig_topo}
\end{figure}

We consider a secure downlink IoT communication scenario as depicted in Fig. \ref{fig_topo}, in which an AP transmits confidential information to a trusted actuator while there exists an eavesdropper aiming to intercept the ongoing transmission.
We assume that the actuator has a single antenna as expected in those resource-constrained IoT applications, and the eavesdropper is equipped with $ K_E $ antennas for strong eavesdropping.
The case in which the AP is equipped with a single antenna will be discussed in Sections \ref{sec_single_performance} and \ref{sec_single_opt}, and the scenario with a multi-antenna AP will be investigated in Section \ref{sec_multi}.

For many IoT applications, the information transmitted from the AP to the actuator is in short packets so as to achieve a low latency (short delay).
Under this case, the blocklength of the channel coding is also finite. Specifically, the AP transmits $ B $ information bits over $ N $ channel uses for each sporadic short-packet transmission.
The channels from the AP to the actuator and the eavesdropper are assumed to undergo independent quasi-static Rayleigh fading combined with large-scale path loss. The fading process stays constant over one transmission block ($ N $ channel uses) while it is independent and identically distributed (i.i.d.) among different blocks.
%As shown in Fig. \ref{fig_topo}, the two channels are denoted by $ h_X=d_X^{-\alpha/2}g_X $ where $ X\in \{A,E\} $, $ d_X $ denotes the distance from the AP, $ \alpha $ denotes the path-loss exponent, and $ g_X\sim \mathcal{CN}(0,1) $ accounts for the small-scale Rayleigh fading.
Moreover, all the channels in the system are corrupted by additive white Gaussian noise (AWGN) in addition to fading and path loss.

\subsection{Achievable Secrecy Rate under Finite Blocklength}\label{subsec_rate}
It is well-known that as long as the transmission rate is below the secrecy capacity and the codeword mapped from the confidential message is sufficiently long, both the error probability and information leakage can be made as small as desired.
However, when the transmission blocklength is finite, decoding error and information leakage at the receiver are inevitable. %\cite{Yang2016d,Yang2017d}.

% sufficiently large value of
According to \cite{Yang2016d} and \cite{Yang2017d}, for a given blocklength $ N $, a constraint on the decoding error probability of $ \epsilon $ at the legitimate receiver, and a secrecy constraint on the information leakage of $ \delta $ measured by the total variation distance\footnote{Its formal mathematical definition was given in \cite[eq. (8)]{Yang2016d}, which describes the statistical independence between the transmitted confidential message and the eavesdropper's observation.}, a lower bound on the maximal secret communication rate (an achievable secrecy rate) can be approximated as follows:\footnote{The equation in \cite[eq. (40)]{Yang2016d} is adapted as here for the complex-valued channel by noting that the equivalent blocklength gets doubled compared with the real-valued one with the same SNR.}
\begin{align}\label{R}
R(N,\epsilon,\delta)=C_S-\sqrt{\frac{V_A}{N}}\frac{Q^{-1}(\epsilon)}{\ln 2}-\sqrt{\frac{V_E}{N}}\frac{Q^{-1}(\delta)}{\ln 2},
\end{align}
conditioned on $ \gamma_A>\gamma_E $ (otherwise the secrecy rate is zero), where $ \gamma_A $ and $ \gamma_E $ are the signal-to-noise ratios (SNRs) at the actuator and the eavesdropper, respectively.
In \eqref{R}, $ C_S=\log_2(1+\gamma_A)-\log_2(1+\gamma_E) $ denotes the secrecy capacity, which characterizes the maximal secrecy rate at which information can be securely and reliably transmitted with codewords of infinite blocklength; $ V_X=1-(1+\gamma_X)^{-2},~X\in \{A,E\} $, is the channel dispersion which measures the stochastic variability of the channel relative to a deterministic channel with the same capacity\cite{Polyanskiy2010};
$ Q^{-1}(\cdot) $ is the inverse of the Gaussian Q-function $ Q(x)=\int_{x}^{\infty}\frac{1}{\sqrt{2\pi}}e^{-\frac{t^2}{2}}dt $.
%TODO AWGN->quasi-static fading channel

The expression in \eqref{R} implies that compared with the secrecy capacity, realizing the targeted decoding error probability $\epsilon$ and satisfying the secrecy constraint on the information leakage $\delta$ respectively incur a penalty term on the achievable secrecy rate with finite blocklength coding, which is proportional to $1/\sqrt{N}$.
When the blocklength $N$ approaches infinity, the penalty diminishes and the achievable secrecy rate given in \eqref{R} asymptotically coincides with the secrecy capacity $C_S$, as expected.
Moreover, according to \eqref{R}, the achievable secrecy rate decreases as the reliability and secrecy constraints become more stringent, i.e., smaller $\epsilon$ and $\delta$.

\subsection{Performance Metrics and Problem Formulation}\label{subsec_metric}
In this work, we use secrecy throughput, i.e., the average secrecy rate at which the data packet is reliably transmitted under a certain secrecy constraint, as the main metric to analyze the performance of systems with secure short-packet communications.
Recalling that the AP transmits $ B $ information bits in total for each sporadic short-packet transmission, its transmission rate is then given by $ R=B/N $ once the blocklength $ N $ is determined.
By substituting $ R=B/N $ into \eqref{R}, the corresponding decoding error probability at the actuator under the tolerance of information leakage $ \delta $ is characterized by
\begin{align}\label{eps}
\epsilon = Q\left(\sqrt{\frac{N}{V_A}}\left(\ln\frac{1+\gamma_A}{1+\gamma_E}-\sqrt{\frac{V_E}{N}}Q^{-1}(\delta)-\frac{B}{N} \ln 2\right)\right),
\end{align}
for $\gamma_A> \gamma_E$.
When $ \gamma_A\leq \gamma_E $, the secrecy capacity is zero and we simply set $ \epsilon=1 $ for this trivial case.
%TODO set eps to 1 when eps>0.5 ?
The average achievable secrecy throughput measured in bits per channel use (BPCU) is then given by
	\begin{align}\label{T}
	T=\E_{\gamma_A,\gamma_E} \left[\frac{B}{N} (1-\epsilon)\right]
	= \frac{B}{N} (1-\bar{\epsilon}),
	\end{align}
	where $ \bar{\epsilon} \triangleq \E_{\gamma_A,\gamma_E} \left[\epsilon\right] $ denotes the average decoding error probability.

Throughout this paper, we assume that the tolerance of information leakage is fixed as $ \delta\in (0,1/2) $ to impose a stringent secrecy constraint.
For a fixed message size of $ B $ bits, the blocklength $ N $ actually captures the corresponding physical-layer transmission latency measured in channel uses\cite{Lopez2017,Yu2018}.
It can be easily seen from \eqref{eps} that the decoding error probability  decreases as $ N $ increases, while a larger $ N $ reduces the transmission data rate and incurs a longer latency. Therefore, it is important to investigate the impact of $ N $ on the secrecy throughput defined in \eqref{T}, which strikes a fundamental tradeoff between transmission latency and decoding error.
Furthermore, there is a similar rate-reliability tradeoff for the total number of transmitted bits $B$, since the decoding error probability is an increasing function of $B$ from \eqref{eps}.
The values of $N$ and $B$ should be carefully chosen so as to maximize the average secrecy throughput.

\section{Analysis for the Single-Antenna Case} \label{sec_single_performance}
In this section, the secrecy throughput for the case with a single-antenna AP under finite blocklength is investigated. We first find a closed-form approximation for the secrecy throughput in Section \ref{subsec_T_approx}, and then some insight is obtained by carefully studying the secrecy throughput in the high-SNR regime as well as in the case with infinite blocklength in Sections \ref{subsec_high_SNR} and \ref{subsec_large_N}, respectively.
%then the behavior of the approximation in the high-SNR and large-blocklength regimes is analyzed in Sections \ref{subsec_high_SNR} and \ref{subsec_large_N}, respectively.

\subsection{Secrecy Throughput Approximation}\label{subsec_T_approx}
For the single-antenna case, the channels from the AP to the actuator and the eavesdropper are represented by $ h_A=d_A^{-\alpha/2}g_A$ and $ \mb{h}_E=d_E^{-\alpha/2}\mb{g}_E$, respectively, where $ d_X$ for $X\in \{A,E\} $ denotes the distance from the AP, $ \alpha $ denotes the large-scale path-loss exponent, and $ g_A\sim \mathcal{CN}(0,1) $ and $ \mb{g}_E\sim \mathcal{CN}(\mb{0},\mb{I}_{K_E}) $ account for the small-scale Rayleigh fading.
The SNRs at the actuator and the eavesdropper are then characterized by
\begin{subequations}\label{gamma}
	\begin{equation}
	\gamma_A = \frac{P|h_A|^2}{\sigma_A^2}=\rho_A |g_A|^2,
	\end{equation}
and
	\begin{equation}
	\gamma_E = \frac{P\|\mb{h}_E\|^2}{\sigma_E^2}=\rho_E \|\mb{g}_E\|^2,
	\end{equation}
\end{subequations}
respectively, where $ P $ is the transmit power at the AP. For $ X\in \{A,E\} $, $ \sigma_X^2 $ is the power of the AWGN at the receiver, and $ \rho_X\triangleq P d_X^{-\alpha}/\sigma_X^2  $.
From \eqref{gamma}, we have $ \gamma_A\sim\mathrm{Exp}(1/\rho_A) $ and $  \gamma_E\sim\mathrm{Gamma}(K_E,\rho_E)$.
The secrecy throughput in \eqref{T} can thereby be calculated as follows:
\begin{align}\label{T1}
T=\frac{B}{N} \int_{0}^{\infty}
\Psi(y)
\frac{y^{K_E-1}e^{-y/\rho_E}}{\rho_E^{K_E}\Gamma(K_E)} dy,
\end{align}
where
\begin{align}\label{Psi}
\Psi(y)\triangleq
\int_{y}^{\infty} \left(1-\epsilon_{\gamma_A|\gamma_E=y}(x) \right)\frac{1}{\rho_A}e^{-\frac{x}{\rho_A}}dx
\end{align}
and $ \epsilon_{\gamma_A|\gamma_E=y}(\cdot) $ is the decoding error probability in \eqref{eps} with respect to (w.r.t.) $ \gamma_A $ conditioned on $ \gamma_E=y $.
Note that the lower limit of the integral $ \Psi(y) $ is $ y $ since we have set $ \epsilon=1 $ when $ \gamma_A\leq \gamma_E $.

To proceed with \eqref{T1}, the main obstacle lies in the calculation of the integral \eqref{Psi}, where the function $ \epsilon_{\gamma_A|\gamma_E}(\cdot) $ has an intractable form.
To circumvent this problem, we first give an approximation of $ \epsilon_{\gamma_A|\gamma_E}(\cdot) $ in the following lemma.

\begin{lemma}\label{lemma_eps}
	A first-order approximation for $ \epsilon_{\gamma_A|\gamma_E}(x) $ is given as follows:
	\begin{align}\label{P}
	\epsilon_{\gamma_A|\gamma_E}(x) &\approx P_{\gamma_A|\gamma_E}(x) \notag\\
	&\triangleq
	\begin{cases}
	1, &x<\frac{1}{2k}+x_0,\\
	\frac{1}{2}+k(x-x_0), &x\in \left[\frac{1}{2k}+x_0,-\frac{1}{2k}+x_0\right],\\
	0,&x>-\frac{1}{2k}+x_0,
	\end{cases}
	\end{align}
	where
	\begin{align}\label{x0}
	x_0\triangleq e^{\sqrt{\frac{V_E}{N}}Q^{-1}(\delta)+\frac{B}{N} \ln 2}(1+\gamma_E)-1,
	\end{align}
    and
	\begin{align}\label{k}
	k\triangleq \frac{d \epsilon_{\gamma_A|\gamma_E}(x)}{dx}\Bigg|_{x=x_0}
	=-\sqrt{\frac{N}{2\pi x_0(x_0+2)}}.
	\end{align}
\end{lemma}
\begin{IEEEproof}
	This approximation follows from the linearization technique used in \cite{Makki2014} and \cite{Makki2016} which have not taken the secrecy constraint into account.
\end{IEEEproof}

\begin{figure}[t]
	\centering
	\includegraphics[width=3in]{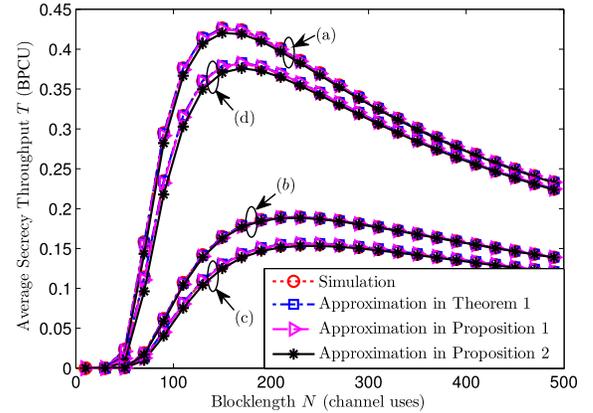}
	\caption{The average secrecy throughput $ T $ and its approximations under the single-antenna case versus blocklength $ N $ with system parameters: (a) $ K_E=2 $, $ \rho_A=10 $ dB, and $ \delta=10^{-2} $; (b) $ K_E=4 $, $ \rho_A=10 $ dB, and $ \delta=10^{-2} $; (c) $ K_E=2 $, $ \rho_A=6 $ dB, and $ \delta=10^{-2} $; (d) $ K_E=2 $, $ \rho_A=10 $ dB, and $ \delta=10^{-4} $. The total transmitted bit number is $ B=200 $ bits, $ \rho_E=3 $ dB, $ M_1=10 $, and $ M_2=20 $.}
	\label{fig_T_single}
\end{figure}

With the approximation in \eqref{P}, the difficulty now lies in
%The integral $ \Psi(y) $ in \eqref{Psi} is still intractable for calculation after leveraging the approximation in Lemma \ref{lemma_eps} due to
the fact that the lower limit of the integral $ \Psi(y) $ is not a constant.
To further simplify $ \Psi(y) $, we approximate it by changing the lower limit of the integral from $ y $ to $ 0 $ based on the fact that $ \epsilon_{\gamma_A|\gamma_E=y}(x)>1/2 $ when $ x<y $.
% and $ \delta<0.5 $.\footnote{The condition $ \delta<0.5 $ usually holds since the leaked secrecy information to the eavesdropper is expected to be little.}
Moreover, according to \eqref{eps} we know that $ \epsilon_{\gamma_A|\gamma_E=y}(x)\to 1 $ when $ N\to \infty $ and $ x<y $, while $ \epsilon \to 1 $ as $ N\to 0 $.
Therefore, this approximation is expected to be tight for a wide range of $ N $, which will be validated by the simulations shown in Fig. \ref{fig_T_single}.
With the aid of the above approximations, the following theorem provides an approximated secrecy throughput.

\begin{theorem}\label{theorem_T}
	The secrecy throughput in \eqref{T1} can be approximated as follows:
	\begin{align}\label{app_T}
	T\approx \frac{B}{N} \int_{0}^{\infty}
	k\rho_A e^{-\frac{x_0}{\rho_A}}\left(e^{\frac{1}{2k\rho_A}}-e^{-\frac{1}{2k\rho_A}}\right) \bigg|_{\gamma_E=y}
	\frac{y^{K_E-1}e^{-y/\rho_E}}{\rho_E^{K_E}\Gamma(K_E)}dy.
	\end{align}
\end{theorem}
\begin{IEEEproof}
	By leveraging Lemma \ref{lemma_eps} and partial integration, the integral $ \Psi(\gamma_E) $ can be approximated as follows:
	\begin{align}\label{app_Psi}
	\Psi(\gamma_E)
	&\approx
	1-\int_{0}^{\infty} P_{\gamma_A|\gamma_E}(x) d\left(1-e^{-\frac{x}{\rho_A}}\right) \notag\\
	&=1+k \int_{\frac{1}{2k}+x_0}^{-\frac{1}{2k}+x_0} \left(1-e^{-\frac{x}{\rho_A}}\right) dx \notag\\
	&=k\rho_A e^{-\frac{x_0}{\rho_A}}\left(e^{\frac{1}{2k\rho_A}}-e^{-\frac{1}{2k\rho_A}}\right).  %\bigg|_{\gamma_E=y}
	%&=k\rho_A e^{-\frac{1}{\rho_A}\left(\frac{1}{2k}+x_0\right)}\left(e^{\frac{1}{k\rho_A}}-1\right)
	\end{align}
	The secrecy throughput can then be approximated by substituting \eqref{app_Psi} into \eqref{T1}.
\end{IEEEproof}

%In many IoT applications, we usually have $10^2\leq N\leq 10^3$ channel uses so as to transmit the data packet with hundreds of bits for most of latency-critical IoT applications according to \cite{Schulz2017}.
Note that the analytical result provided in Theorem \ref{theorem_T} is applicable to the case with arbitrary blocklength, but its form is quite complicated. In order to obtain more insight about the secrecy throughput, some approximations in the moderate-blocklength regime, i.e., $10^2\leq N\leq 10^3$, will be carried out. It is important to point out that the case with $10^2\leq N\leq 10^3$, i.e., a data packet containing hundreds of bits, is particularly important to latency-critical IoT applications as shown in \cite{Schulz2017}.
Therefore, in the moderate-blocklength regime the parameter $ |k| $ can be large based on the fact that $ |k| $ is an increasing function of $N$ according to \eqref{k}, and we further have $ e^{\pm\frac{1}{2k\rho_A}}\approx 1\pm\frac{1}{2k\rho_A} $ especially when $ \rho_A $ is large.
%or the actuator works in the high SNR regime, we have $ e^{\frac{1}{2k\rho_A}}-e^{-\frac{1}{2k\rho_A}}\approx \frac{1}{k\rho_A} $.
Accordingly, when $|k\rho_A|$ is large (the relative error for the approximation of $ \Psi(\gamma_E) $ is smaller than 5\% when $|k\rho_A|>1$), $ \Psi(\gamma_E) $ in \eqref{app_Psi} can be further approximated as follows:
\begin{align}\label{app_Psi2}
\Psi(\gamma_E) \approx e^{-\frac{x_0}{\rho_A}}.
\end{align}
By leveraging \eqref{app_Psi2}, a closed-form approximation for the secrecy throughput of the system with a single-antenna AP is then given in the following proposition.

\begin{proposition}\label{prop_T}
	When $|k\rho_A|$ is large, %$ \rho_A $ is large and $10^2\leq N\leq 10^3$,
	 the secrecy throughput of the system with a single-antenna AP under finite blocklength can be approximated as follows:
	\begin{align}\label{app_T1}
	T\approx \frac{B}{N \rho_E^{K_E}\Gamma(K_E)}
	\Bigg(&
	\frac{M_1}{2} \sum_{n=1}^{M_2}\left(\varrho_{M_2} f\left(\frac{M_1}{2}(t_n+1)\right) \sqrt{1-t_n^2} \right) \notag\\
	&\quad +\frac{1}{\varpi_2^{K_E}} e^{-\frac{\varpi_1-1}{\rho_A}}\Gamma(K_E,\varpi_2 M_1)\Bigg),
	\end{align}
	where $ M_1 $ is a sufficiently large parameter to ensure $ V_E\approx 1 $ when $ \gamma_E>M_1 $, $ M_2 $ is a parameter for the complexity-accuracy tradeoff, $ \varrho_{M_2} \triangleq \frac{\pi}{M_2} $, $ f(z) \triangleq z^{K_E-1} e^{-\left(\frac{x_0(z)}{\rho_A}+\frac{z}{\rho_E}\right)} $ with $ x_0(z)\triangleq x_0|_{\gamma_E=z} $, $ t_n\triangleq \cos\left(\frac{2n-1}{2M_2}\pi \right) $, $ \varpi_1\triangleq e^{\frac{Q^{-1}(\delta)}{\sqrt{N}}+\frac{B}{N} \ln 2} $, and $ \varpi_2\triangleq \frac{\varpi_1}{\rho_A}+\frac{1}{\rho_E} $.
\end{proposition}
\begin{IEEEproof}
	The proof is provided in Appendix \ref{app_prop_T_single}.
\end{IEEEproof}

%As validated by the simulations in Fig. \ref{fig_T_single}, the approximation given in Proposition \ref{prop_T} matches well with the simulation results under a wide range of system parameters. However,

\begin{table}[t]
	\centering
	\caption{Proposed Analytical Framework \protect\\ for Analyzing the Secrecy Throughput}
	%\vspace{-3mm}
	\label{tab_framework}
	\begin{tabular}{|p{0.0\columnwidth}p{0.89\columnwidth}|}
		\hline
		\hline
		1.&Formulate the secrecy throughput according the definition given in \eqref{T};\\
		2.&For any given eavesdropper's SNR $ \gamma_E $, approximate the decoding error probability $ \epsilon_{\gamma_A|\gamma_E}(x) \approx P_{\gamma_A|\gamma_E}(x) $ w.r.t. the actuator's SNR $ \gamma_A $ according to Lemma \ref{lemma_eps};\\
		3.&Approximate the integral $ \Psi(\gamma_E) $ by changing the lower limit of the integral from $ y $ to $ 0 $, and then the closed-form expression of $ \Psi(\gamma_E)  $ can be directly calculated (in the single-antenna AP case) or further approximated by using the first-order Riemann integral (in the multi-antenna AP case); \\
		4.&By substituting the result of $ \Psi(\gamma_E)  $ into the expression of the secrecy throughput, the closed-form expression of $ T $ can be further obtained by splitting the integral and using Gaussian-Chebyshev quadrature. \\
		\hline\hline
	\end{tabular}
	%\vspace{-5mm}
\end{table}

Our proposed analytical framework for analyzing the secrecy throughput and obtaining a closed-form expression for short-packet communications is summarized in Table \ref{tab_framework}.
The expression given in Proposition \ref{prop_T} takes a cumbersome form, and it is still difficult to gain any insight on the system performance from \eqref{app_T1}.
%it is still difficult to gain any insight due to the sophisticated form of \eqref{app_T1}.
In what follows, we aim to further simplify the expression of the secrecy throughput so as to analytically investigate the intrinsic latency-reliability tradeoff under the secrecy constraint for the system with finite blocklength coding.

Since the eavesdropper is equipped with multiple antennas and maximal-ratio combining is employed, its SNR can be considerably large especially when the AP has a large transmit power. In this regard, a small parameter $ M_1 $ can be chosen in Proposition \ref{prop_T} without loss of too much accuracy.
%In the high SNR regime ($\rho_A$ and $ \rho_E $ are large), we remark that the integral $ \Omega_1 $
Following this intuition, we provide another approximation for the secrecy throughput in the following proposition, in which a simple form is developed to gain more insight and to conduct optimization for the secrecy system performance with finite blocklength coding.
\begin{proposition}\label{prop_T2}
%	When $10^2\leq N\leq 10^3$, and the average SNR at the eavesdropper $ \E\left[\gamma_E\right] =K_E \rho_E $ and $ \rho_A $ are large,
	When $|k\rho_A|$ and the average SNR at the eavesdropper $ \E\left[\gamma_E\right] =K_E \rho_E $ are large,
	the secrecy throughput of the system with a single-antenna AP under finite blocklength can be further approximated as follows:
	\begin{align}\label{app_T2}
	T\approx \frac{B}{N \left(\rho_E\varpi_2\right)^{K_E}} e^{-\frac{\varpi_1-1}{\rho_A}}.
	\end{align}
\end{proposition}
\begin{IEEEproof}
	The result can be directly obtained by setting $ M_1=0 $ in \eqref{app_T1} and by noting that $ \Gamma(K_E,0)=\Gamma(K_E) $.
\end{IEEEproof}

As depicted in Fig. \ref{fig_T_single}, the approximations provided in Propositions \ref{prop_T} and \ref{prop_T2} match well with the simulations for a wide range of $ N $.
% and system parameters especially when $\rho_A$ and $ K_E\rho_E $ are large.
From Fig. \ref{fig_T_single}, the secrecy throughput is an increasing function of $ \rho_A $ and information leakage $ \delta $, while it decreases with an increase in $ K_E $.
Moreover, there exists an optimal blocklength to maximize the secrecy throughput, which will be analytically studied in Section \ref{sec_single_opt}.
%Relying on the elegant form of the approximation in \eqref{app_T2}, we will theoretically analyze the optimum and gain more insight in what follows.
Because of the accuracy of the results provided in Proposition \ref{prop_T2} and their analytical tractability, we will use \eqref{app_T2} to optimize the system performance hereinafter.

\subsection{High-SNR Regime}\label{subsec_high_SNR}
When the transmit power $ P $ at the AP approaches to infinity, we know that both $ \rho_A $ and $ \E\left[\gamma_E\right] $ approach to infinity.
Under this regime, the approximation in \eqref{app_T2} is accurate
for a wide range of $N$ according to the condition in Proposition \ref{prop_T2}.
%for a wide any $N>0$ based on the fact that the approximation in \eqref{app_Psi2} is accurate since $ |k\rho_A|\to \infty $.
%in the moderate-blocklength regime according to the condition in Proposition \ref{prop_T2}.
Therefore, by letting $ P $ in \eqref{app_T2} go to infinity, the average secrecy throughput in the high-SNR regime is given by
\begin{align}\label{T_inf_P}
T^{P\to \infty}=\frac{B}{N} \left(1-\bar{\epsilon}^{P\to \infty}\right),
%&=\frac{B}{N}
%\left(1+\frac{\sigma_A^2}{\sigma_E^2} \left(\frac{d_A}{d_E}\right)^\alpha \varpi_1 \right)^{-K_E}.
\end{align}
where the average decoding error probability in the high-SNR regime is denoted by
\begin{align}\label{eps_inf_P}
\bar{\epsilon}^{P\to \infty}=1-\left(1+\frac{\sigma_A^2}{\sigma_E^2} \left(\frac{d_A}{d_E}\right)^\alpha \varpi_1 \right)^{-K_E}.
\end{align}
From \eqref{eps_inf_P} we know that the average decoding error probability cannot decrease to zero as the transmit power at the AP approaches infinity, since both the SNRs at the actuator and the eavesdropper increase at the same time.
Moreover, in the high-SNR regime, the secrecy throughput increases with $ d_E $ and $ \delta $, while it is a decreasing function of $ d_A $ and $ K_E $.

\subsection{The Classical Case with Infinite Blocklength}\label{subsec_large_N}
To gain more insight on the asymptotic system performance and understand the connection between the cases with finite and infinite blocklength, in this subsection we focus on the classical case with infinite blocklength, i.e., the blocklength $ N $ approaches to infinity. %\footnote{Note that only this subsection is to consider the case with infinite blocklength, and the remaining parts of this paper are still to focus on the moderate blocklength regime.}
	
When $N\to \infty$, we know that the decoding error probability $ \epsilon\to 0 $ as long as $ \gamma_A>\gamma_E $ (otherwise $ \epsilon\to 1 $) from \eqref{eps}.
Therefore, with infinite blocklength, the secrecy throughput defined in \eqref{T} can be expressed as follows:
\begin{align}\label{T_inf_N}
T^{N\to \infty}
&=\frac{B}{N}\P\{\gamma_A>\gamma_E\}%\notag\\&
=\frac{B}{N} \left(\frac{\rho_A}{\rho_A+\rho_E}\right)^{K_E}.
\end{align}
Note that the result given in \eqref{T_inf_N} can be directly obtained from \eqref{app_T2} by letting $ N\to \infty $ for any positive $ \delta $, which verifies the accuracy of our approximation in Proposition \ref{prop_T2} in the large-blocklength regime.

It is also worth noting that the result in \eqref{T_inf_N} coincides with the case of infinite blocklength, which is expected.
With infinite blocklength and appropriate transmission rates, there is no decoding error once the secrecy capacity $ C_S $ is not zero, namely, $ \gamma_A>\gamma_E $.
Moreover, under the condition $ \gamma_A>\gamma_E $ since the transmission rate $ \frac{B}{N}\to 0 $ which is smaller than $ C_S $ as $ N\to \infty $, the rate $ \frac{B}{N} $ is always achievable without any information leakage to the eavesdropper, i.e., the result in \eqref{T_inf_N} holds for any positive $ \delta $.

\section{Optimization for the Single-Antenna Case} \label{sec_single_opt}
In this section, the system performance under the single-antenna case is maximized by optimizing the blocklength. We first investigate unconstrained secrecy throughput optimization in Section \ref{subsec_opt}, and then throughput optimization is studied with the reliability and latency constraints in Section \ref{subsec_opt_cons}.

\subsection{Unconstrained Secrecy Throughput Optimization}\label{subsec_opt}
As indicated in Section \ref{subsec_metric}, blocklength $ N $ plays an important role in optimizing the secrecy throughput, where a balanced tradeoff between transmission latency and decoding error needs to be realized.
In particular, we focus on the question how to choose blocklength $ N $ for secrecy throughput maximization when the message size is fixed to be $ B $ bits in total.
The following theorem characterizes the optimal blocklength which maximizes the secrecy throughput.
\begin{theorem}\label{theorem_N}
	The secrecy throughput $ T $ in \eqref{app_T2} is a quasi-concave function of the relaxed continuous blocklength $ N $.
	The optimal blocklength can be chosen as the one from $ \left\{\lceil N^* \rceil, \lfloor N^* \rfloor\right\}  $ that yields the largest secrecy throughput, where $ N^*>0 $ is the unique root of
	\begin{align}\label{N}
	\Xi(N)\triangleq \frac{\varpi_1}{\rho_A}\left(\frac{Q^{-1}(\delta)}{2\sqrt{N}}+\frac{B}{N} \ln 2\right)\left(1+\frac{K_E}{\varpi_2}\right)-1=0.
	\end{align}
\end{theorem}
\begin{IEEEproof}
	The proof is provided in Appendix \ref{app_theorem_N}.
\end{IEEEproof}

According to Theorem \ref{theorem_N}, the optimal blocklength can be obtained through a bisection search.
With the aid of the relation in \eqref{N}, insight on the behavior of $ N^* $ is developed in the following corollary.
\begin{corollary}\label{corollary_N}
	The optimal $ N^* $ increases with $ \rho_E $, $ K_E $, and $ B $, while it decreases as $ P $, $ \rho_A $, and $ \delta<0.5 $ increase.
\end{corollary}
\begin{IEEEproof}
	The proof is provided in Appendix \ref{app_corollary_N}.
\end{IEEEproof}

From Corollary \ref{corollary_N}, we learn that the decoding error becomes more pronounced compared with transmission latency when the average SNR at the eavesdropper $ K_E\rho_E $ or the transmission bit number $ B $ gets larger. Thus, the optimal blocklength $ N^* $ becomes larger accordingly, which aligns with our intuition.
Contrarily, when either the transmit power $ P $ at the AP, the SNR at the actuator $ \rho_A $, or the tolerance of the information leakage $ \delta $ increases, the probability of decoding error decreases and the optimal blocklength $ N^* $ becomes smaller in order to reduce transmission latency.

After obtaining the optimal blocklength according to Theorem \ref{theorem_N}, the optimal secrecy throughput denoted by $ T(N^*) $ can be evaluated by substituting $ N^* $ in \eqref{app_T2}.
The following corollary provides the impact of the related system parameters on $ T(N^*) $.
\begin{corollary}\label{corollary_T}
	The optimal $ T(N^*) $ decreases with an increase in $ \rho_E $ or $ K_E $, while it increases with $ B $, $ \delta$, $ P $, and $ \rho_A $.
\end{corollary}
\begin{IEEEproof}
	The proof is provided in Appendix \ref{app_corollary_T}.
\end{IEEEproof}

According to the partial derivative of the secrecy throughput $ T $ w.r.t. the total number of transmitted bits $ B $ given by \eqref{T_B} in the proof of Corollary \ref{corollary_T}, we learn that $ T $ is also a quasi-concave function of $ B $.
Therefore, a similar conclusion can be drawn when we optimize $ B $ to maximize $ T $ for a fixed $ N $ as in Theorem \ref{theorem_N}.
Furthermore, it is interesting to see that from Corollary \ref{corollary_T}, the optimal $ T(N^*) $ is a monotonically increasing function of $ B $ instead of being quasi-concave once the blocklength is chosen as the optimal one.
An illustration of the impact of these system parameters will be shown in Fig. \ref{fig_T_B_optN} in Section \ref{subsec_sim_single}.

So far, we have investigated the optimization of the secrecy throughput without any reliability and latency constraints, i.e., for any average decoding error probability and blocklength. In the next subsection, we will focus on the scenario with these practical constraints.

\subsection{Optimization Under Reliability and Latency Constraints}\label{subsec_opt_cons}
For some short-packet communication applications such as remote control and industrial automation, high-reliability and low-latency communication is favorable, which aligns with the requirements of uRLLC.
In this subsection, we focus on the secrecy throughput optimization under these application scenarios.

Mathematically, we aim to solve the following optimization problem
\begin{subequations}\label{p0}
	\begin{align}
	\maxp_{N\in \mathbb{N}_+} \quad &T \\
	\st \quad &\bar{\epsilon}\leq \zeta_\epsilon, \label{cons_eps}\\
	&N\leq \zeta_N, \label{cons_N}
	\end{align}
\end{subequations}
where \eqref{cons_eps} accounts for the decoding reliability constraint and $ \zeta_\epsilon $ is the maximum tolerance for the average decoding error probability $ \bar{\epsilon} $, while \eqref{cons_N} captures the constraint of physical-layer transmission latency by imposing a maximum tolerable blocklength $ \zeta_N $.
Note that to make problem \eqref{p0} tractable, the other latency parameters, such as the propagation latency and the processing time for encoding and decoding, are assumed to be fixed and have been absorbed into parameter $ \zeta_N $.

With the approximation given in Proposition \ref{prop_T2}, we have $ \bar{\epsilon}\approx 1-\frac{1}{(\rho_E\varpi_2)^{K_E}} e^{-\frac{\varpi_1-1}{\rho_A}} $, which is a decreasing function of $ N $.
Therefore, problem \eqref{p0} is then recast as follows:
 \begin{subequations}\label{p1}
 	\begin{align}
 	\maxp_{N\in \mathbb{N}_+} \quad &T \\
 	\st \quad &\bar{\epsilon}^{-1} (\zeta_\epsilon) \leq N\leq \zeta_N,
 	\end{align}
 \end{subequations}
where $ \bar{\epsilon}^{-1}(\cdot) $ is the inverse function of $ \bar{\epsilon}(N) $ with $ \bar{\epsilon}(N) $ being the function of $ \bar{\epsilon}$ w.r.t. $ N $.
Note that the solution to problem \eqref{p1} exists only when $ \left\lceil\bar{\epsilon}^{-1} (\zeta_\epsilon)\right\rceil \leq \left\lfloor\zeta_N\right\rfloor $.
According to the conclusion given in Theorem \ref{theorem_N}, the following corollary shows the optimal blocklength for problem \eqref{p1}.
\begin{corollary}\label{theorem_N1}
	When $ \left\lceil\bar{\epsilon}^{-1} (\zeta_\epsilon)\right\rceil \leq \left\lfloor\zeta_N\right\rfloor $, the optimal blocklength for problem \eqref{p1} is given by
	\begin{align}\label{N1}
	N^\#=
	\begin{cases}
	\left\lceil\bar{\epsilon}^{-1} (\zeta_\epsilon)\right\rceil ,& N^*\leq	\left\lceil\bar{\epsilon}^{-1} (\zeta_\epsilon)\right\rceil,\\
	\arg\max_{N\in \left\{\lceil N^* \rceil,\lfloor N^* \rfloor\right\}}~T(N), & \left\lceil\bar{\epsilon}^{-1} (\zeta_\epsilon)\right\rceil < N^* < \left\lfloor\zeta_N\right\rfloor,\\
	\left\lfloor\zeta_N\right\rfloor,&N^* \geq \left\lfloor\zeta_N\right\rfloor,
	\end{cases}
	\end{align}
	where $ N^* $ is defined in Theorem \ref{theorem_N}.
\end{corollary}
\begin{IEEEproof}
	We first relax $ N $ in problem \eqref{p1} as a continuous positive value. Then the result directly follows from the fact that the secrecy throughput $ T $  is a quasi-concave function of the continuous $ N $ according to Theorem \ref{theorem_N}.
\end{IEEEproof}

The optimal average secrecy throughput obtained from Corollary \ref{theorem_N1} under the reliability and latency constraints will be shown in Fig. \ref{fig_opt_T_cons} in Section \ref{subsec_sim_single}.
When jointly optimizing the blocklength $ N $ and transmitted bit number $ B $ for secrecy throughput maximization, we remark that Corollary \ref{theorem_N1} can be easily expanded by conducting a one-dimensional search for $ B $ and then finding the optimal $ N $ for each fixed $ B $ according to \eqref{N1}.

\section{Analysis for the Multi-Antenna Case} \label{sec_multi}
In this section, we focus on the secrecy throughput with a multi-antenna AP under finite blocklength.
Compared with the single-antenna case, the transmission security of the system can be greatly improved when the AP has multiple antennas since the extra spatial degrees of freedom can be leveraged.

We consider a scenario where the AP is equipped with $ K_A $ antennas.
As in the single-antenna case, we assume that the AP has the statistical CSI of the eavesdropper only due to its passive nature.\footnote{Compared with the full CSI, statistical CSI is more practical to be obtained.  This is because statistical CSI depends heavily on the propagation environment and  is expected to vary more slowly with time than instantaneous CSI, which can be acquired in practical communication scenarios. This is also a common assumption in the literature of physical-layer security \cite{Mukherjee2014,Mukherjee2015,Yang2015a,Liu2017d,Poor2017}.}
The channels from the AP to the actuator and the eavesdropper are similarly defined as $ \mb{h}_A^T=d_A^{-\alpha/2}\mb{g}_A^T $ and $ \mb{H}_E=d_E^{-\alpha/2}\mb{G}_E $, respectively, where each element in $ \mb{g}_A^T \in \C^{1\times K_A}$ and $ \mb{G}_E\in\C^{K_E\times K_A} $ follows an i.i.d. $\mathcal{CN}(0,1) $ distribution.

To degrade the performance of the eavesdropper while ensuring the reliability of the signal reception at the actuator, the AP can leverage either the maximal-ratio transmission (MRT) beamforming scheme or an AN-aided transmission scheme as in \cite{Goel2008,Zheng2015a}.
Specifically, in the AN-aided transmission scheme, the AP uses MRT to transmit confidential signal and injects AN in the nullspace of the legitimate communication channel to confuse the eavesdropper.
Accordingly, the designed transmit signal under the two transmission schemes can be written as the following unified expression:
\begin{align}\label{x}
\mb{x}=\sqrt{\eta P} \mb{w} s + \sqrt{(1-\eta)P}\mb{U}\mb{v},
\end{align}
where $ s $ is the information-bearing signal with unit power, $ \mb{w}\triangleq \mb{g}_A/\|\mb{g}_A\| $ is the MRT beamforming vector, $ \mb{v}\sim \mathcal{CN}(\mb{0},\frac{1}{K_A-1}\mb{I}_{K_A-1}) $ is the injected $ (K_A-1)\times 1 $ AN vector to confuse the eavesdropper, and $ \eta $ is the power allocation factor. The columns of $ \mb{U}\in \C^{K_A\times (K_A-1)} $ span the nullspace of $ \mb{h}_A^T $, and the columns of $ \mb{W}\triangleq [\mb{w},\mb{U}] $ forms an orthogonal basis.
Note that the AN-aided transmission scheme reduces to the MRT beamforming one when $ \eta=1 $ is set in \eqref{x}.

Given the transmit signal in \eqref{x}, the signal-to-interference-plus-noise ratios (SINR) at the actuator is represented as follows:
\begin{align}
&\gamma_A=\frac{\eta P \|\mb{h}_A\|^2}{\sigma_A^2}=\eta\rho_A \|\mb{g}_A\|^2.
\end{align}
As for the eavesdropper, its received signal is given by
\begin{align}
\mb{y}_E=\sqrt{\eta P} \mb{H}_E\mb{w} s + \sqrt{(1-\eta)P}\mb{H}_E\mb{U}\mb{v} + \mb{n}_E,
\end{align}
where $ \mb{n}_E\sim \mathcal{CN}(\mb{0},\sigma_E^2\mb{I}_{K_E}) $ accounts for the AWGN.
By employing the minimum mean-squared error (MMSE) receiver, the corresponding SINR at the eavesdropper is expressed as follows:
\begin{align}
&\gamma_E \notag\\
&=\eta P \mb{w}^H\mb{H}_E^H \left(\frac{(1-\eta)P}{K_A-1}\mb{H}_E\mb{U} \mb{U}^H\mb{H}_E^H+\sigma_E^2\mb{I}_{K_E}\right)^{-1} \mb{H}_E\mb{w}\notag\\
&=\eta \mb{w}^H\mb{G}_E^H \left(\frac{1-\eta}{K_A-1}\mb{G}_E\mb{U} \mb{U}^H\mb{G}_E^H+\frac{1}{\rho_E}\mb{I}_{K_E}\right)^{-1} \mb{G}_E\mb{w}.
\end{align}

Following steps similar to the ones for the single-antenna case, by leveraging Lemma \ref{lemma_eps} and partial integration, we first approximate the integral $ \Psi(\gamma_E) $ as follows:
\begin{align}\label{app_Psi_m}
\Psi(\gamma_E)
&\approx
%1-\int_{0}^{\infty} P_{\gamma_A|\gamma_E}(x) d F_{\gamma_A}(x) \notag\\
%&=
1+k \int_{\frac{1}{2k}+x_0}^{-\frac{1}{2k}+x_0} F_{\gamma_A}(x) dx,
\end{align}
where $ F_{\gamma_A}(x)=\frac{1}{\Gamma(K_A)}\gamma(K_A,\frac{x}{\eta\rho_A}) $ represents the cumulative distribution function (CDF) of $ \gamma_A $ due to $ \|\mb{g}_A\|^2 \sim\mathrm{Gamma}(K_A,1) $.
By noting that
%blocklength $ N $ cannot be set too small for decoding error mitigation
%we usually have $10^2\leq N\leq 10^3$ channel uses
%and parameter $ |k| $ is thereby large from \eqref{k},
in the moderate-blocklength regime the parameter $ |k| $ can be large based on the fact that $ |k| $ is an increasing function of $N$ according to \eqref{k},
the first-order Riemann integral approximation $ \int_{a}^{b}f(x)dx\approx (b-a)f(\frac{a+b}{2}) $ is expected to be tight.
Following this, we further approximate \eqref{app_Psi_m} as follows:
\begin{align}
\Psi(\gamma_E)
&\approx
1- F_{\gamma_A}(x_0).
%\overset{(b)}{=}e^{-\frac{x}{\eta\rho_A}}\sum_{n=0}^{K_A-1}\frac{x^n}{(\eta\rho_A)^n n!}.
\end{align}
Then the secrecy throughput in the multi-antenna case can be calculated according to
\begin{align}\label{T_m}
T=& \frac{B}{N} \int_{0}^{\infty}
\Psi(y) d F_{\gamma_E}(y),
\end{align}
where $ F_{\gamma_E}(\cdot) $ is the CDF of the random variable $ \gamma_E $.
Its specific expression for any antenna number at the eavesdropper $ K_E $ is provided in the following lemma.
\begin{lemma}\label{lemma_gamma_E}
	The CDF of the SINR at the eavesdropper $ \gamma_E $ under the AN-aided transmission scheme can be found as follows:
	\begin{align}\label{F_gamma_E}
	F_{\gamma_E}(x)=1-e^{-\frac{1}{\eta\rho_E}x} \sum_{n=1}^{K_E} \frac{A_n(x)}{(n-1)!}\left(\frac{x}{\eta\rho_E}\right)^{n-1},
	\end{align}
	where
	\begin{align}\label{An}
	A_n(x)\triangleq
	\begin{cases}
		1,&K_E \geq K_A-1+n,\\
		\frac{\sum_{m=0}^{K_E-n}\binom{K_A-1}{m}(\tau x)^m}{(1+\tau x)^{K_A-1}},&K_E < K_A-1+n,
	\end{cases}
	\end{align}
	 with $ \tau\triangleq \frac{\eta^{-1}-1}{K_A-1}$.
\end{lemma}
\begin{IEEEproof}
	The result directly follows from \cite[eq. (11)]{Gao1998}.
\end{IEEEproof}

With the aid of Lemma \ref{lemma_gamma_E}, an approximation for the secrecy throughput under the multi-antenna case is given in the following theorem.
\begin{theorem}\label{theorem_T_multi}
	The secrecy throughput of the system with a multi-antenna AP under finite blocklength can be approximated as follows:
	\begin{align}\label{app_T_multi}
	T\approx &\frac{B}{N}
	\Bigg(
	\frac{M_1}{2} \sum_{n=1}^{M_2}\left(\varrho_{M_2} g\left(\frac{M_1}{2}(t_n+1)\right) \sqrt{1-t_n^2} \right) \notag\\
	&+
	\frac{1}{\Gamma(K_A)}\Gamma\left(K_A,\frac{\varpi_1(M_1+1)-1}{\eta\rho_A}\right) 
	-\frac{\varpi_1 e^{-\frac{\varpi_1-1}{\eta\rho_A}}}{(\eta\rho_A)^{K_A}\Gamma(K_A)}
	\notag\\
	&\times	
	\sum_{p=0}^{K_A-1}\binom{K_A-1}{p}  \left(\frac{\varpi_1}{\varpi_1-1}\right)^p
	\sum_{n=1}^{K_E} \frac{(\varpi_1-1)^{K_A-1}}{(n-1)!(\eta\rho_E)^{n-1}}
	\Theta_n
	\Bigg),
	\end{align}
	where $ M_1 $ is a sufficiently large parameter to ensure $ V_E\approx 1 $ when $ \gamma_E>M_1 $, $ M_2 $ is a parameter for the complexity-accuracy tradeoff, $ \varrho_{M_2} \triangleq \frac{\pi}{M_2} $,  $ g(z)\triangleq F_{\gamma_E}(z) f_{\gamma_A}\left(x_0(z)\right) x'_0(z) $ with
	\begin{align}
	f_{\gamma_A}(x)=\frac{x^{K_A-1}e^{-\frac{x}{\eta\rho_A}}}{(\eta\rho_A)^{K_A}\Gamma(K_A)}
	\end{align}	
	being the probability density function (PDF) of $ \gamma_A $ and
	\begin{align}
	x'_0(y)
	&\triangleq \frac{\partial x_0}{\partial \gamma_E}\Bigg|_{\gamma_E=y}\notag\\
	&=e^{\sqrt{\frac{V_E(y)}{N}}Q^{-1}(\delta)+\frac{B}{N} \ln 2}\left(1+\frac{Q^{-1}(\delta)}{\sqrt{N}(1+y)\sqrt{y(y+2)}}\right)
	\end{align}
	with $ V_E(y)\triangleq  V_E|_{\gamma_E=y}$,
	$ t_n\triangleq \cos\left(\frac{2n-1}{2M_2}\pi \right) $,
	and $ \varpi_3 \triangleq  \frac{\varpi_1 \rho_E+\rho_A}{\eta\rho_A\rho_E}$.
	The calculation of integral $ \Theta_n $ in \eqref{app_T_multi} is discussed as follows:
	
	\emph{1) The case of $ \eta=1 $ (MRT beamforming):}
	\begin{align}
	\Theta_n =\Theta_{1,n}\triangleq
	\frac{1}{\varpi_3^{n+p}}\Gamma(n+p,\varpi_3 M_1),\quad \text{for}~1\leq n \leq K_E.
	\end{align}
	
	\emph{2) The case of $ \eta\neq 1 $ and $ K_E<K_A $:} For $ 1\leq n \leq K_E $,
	\begin{align}
	\Theta_n &=\Theta_{2,n}\notag\\
	&\triangleq
	 \sum_{m=0}^{K_E-n}\binom{K_A-1}{m} \frac{\tau^m e^{-\varpi_3 M_1}}{(1+\tau M_1)^{K_A-1}}\notag\\
	 & \quad \times
	\sum_{q=0}^{p+m+n-1}\binom{p+m+n-1}{q} M_1^{p+m+n-1-q}\left(\frac{1+\tau M_1}{\tau}\right)^{q+1}\notag\\
	& \quad \times
	\Gamma(q+1) U\left(q+1,q-K_A+3,\frac{1+\tau M_1}{\tau} \varpi_3\right),
	\end{align}
	where $ U(\cdot,\cdot,\cdot) $ is the Tricomi's (confluent hypergeometric) function and can be easily evaluated by computational software.
	
	\emph{3) The case of $ \eta\neq 1 $ and $ K_E\geq K_A $:}
	\begin{align}
	\Theta_n =
	\begin{cases}
	\Theta_{1,n},&1\leq n\leq K_E-K_A+1,\\
	\Theta_{2,n},&K_E-K_A+2\leq n\leq K_E.
	\end{cases}
	\end{align}
\end{theorem}
\begin{IEEEproof}
The proof is provided in Appendix \ref{app_theorem_T_multi}.
\end{IEEEproof}

\begin{figure}[t]
	\centering
	\includegraphics[width=3in]{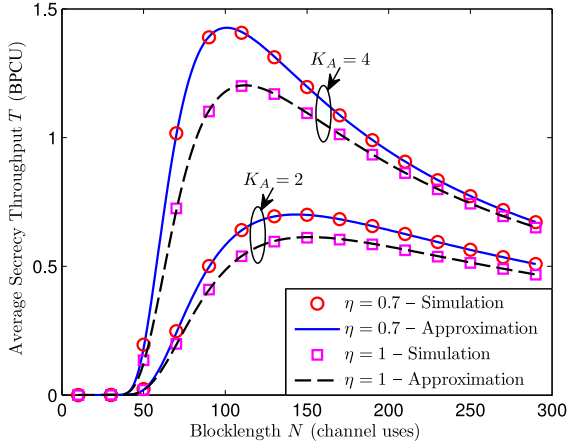}
	\caption{The average secrecy throughput $ T $ and its approximation under the multi-antenna case versus blocklength $ N $ with different antenna numbers at the AP $ K_A $ and power allocation factors $ \eta $. The other system parameters are: $ B=200 $ bits, $ K_E=3 $, $ \rho_A=10 $ dB, $ \rho_E=3 $ dB, $ \delta=10^{-2} $, $ M_1=30 $, and $ M_2=20 $.}
	\label{fig_T_multi}
\end{figure}

We remark that the result presented in Theorem \ref{theorem_T_multi} is general and holds for all possible values of $ \eta $ and $ K_E $ under $ K_A>1 $.
Fig. \ref{fig_T_multi} plots the impact of blocklength on the average secrecy throughput and its approximation in Theorem \ref{theorem_T_multi} with a multi-antenna AP. It can be seen from the figure that the theoretical approximation is close to the simulation result, which verifies the accuracy of the proposed approximations.
Additionally, from Fig. \ref{fig_T_multi} we know that the secrecy throughput increases with the antenna number of the AP.
Most importantly, allocating appropriate power for AN to confuse the eavesdropper can be beneficial for improving the secrecy throughput, meaning that the AN-aided transmission scheme is superior to the plain MRT beamforming scheme by setting an appropriate $\eta$.
Note that since the expression of the secrecy throughput under the case with a multi-antenna AP is intractable to perform further theoretical analysis and optimization, we will comprehensively analyze the system performance through the simulations in Section \ref{subsec_sim_multi}.

\section{Numerical Results}\label{sec_sim}
In this section, we provide numerical results to show performance of the short-packet communication for secure IoT applications.
The parameter settings are as follows, unless otherwise specified:
$ B=200 $, $ \delta=10^{-2} $, $ K_A=2 $, $ K_E=3 $, $ \rho_A=10 $ dB, $ \rho_E=3 $ dB, $ \eta=0.7 $, $ M_1=10 $, and $ M_2=20 $.
%The total transmitted bit number is $ B=200 $ bits, the tolerance of information leakage $ \delta=10^{-2} $, the antenna number at the eavesdropper is $ K_A=2 $, $ K_E=3 $, $ \rho_A=10 $ dB, $ \rho_E=3 $ dB, $ \eta=0.7 $, $ M_1=10 $, and $ M_2=20 $.
All the simulation results shown in this paper are obtained by averaging over $ 100,000 $ channel realizations.

\subsection{Single-Antenna Case}\label{subsec_sim_single}

\begin{figure}[t]
	\centering
	\includegraphics[width=3in]{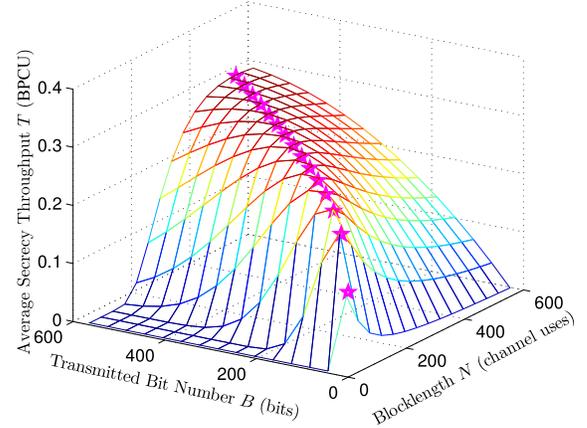}
	\caption{The average secrecy throughput $ T $ obtained by the simulation versus blocklength $ N $ and the transmitted bit number per block $ B $ under the case with a single-antenna AP, where the star markers denote the optimal blocklength obtained by Theorem \ref{theorem_N} and the corresponding approximated secrecy throughput given in Proposition \ref{prop_T2}.}
	\label{fig_T_B_optN}
\end{figure}

Fig. \ref{fig_T_B_optN} depicts the average secrecy throughput obtained by the simulation versus the blocklength $ N $ and the transmitted bit number per block $ B $ under the case with a single-antenna AP. For each given $ B $, the star markers in Fig. \ref{fig_T_B_optN} present the optimal blocklength and secrecy throughput obtained from the proposed analytical approximations.
From Fig. \ref{fig_T_B_optN}, it can be clearly seen that the analytical results coincide well with the simulations, and the method proposed in Theorem \ref{theorem_N} can be used to accurately find the optimal blocklength for secrecy throughput maximization.
Moreover, for each fixed $ B $, it can be observed from Fig. \ref{fig_T_B_optN} that the secrecy throughput is a quasi-concave function of the relaxed continuous blocklength, and there exists an optimal $ N $ which increases with $ B $.
Additionally, the optimal secrecy throughput is also an increasing function of $ B $.

\begin{figure}[t]
	\centering
	\subfloat[]{\label{fig_opt_T}
		\includegraphics[width=3in]{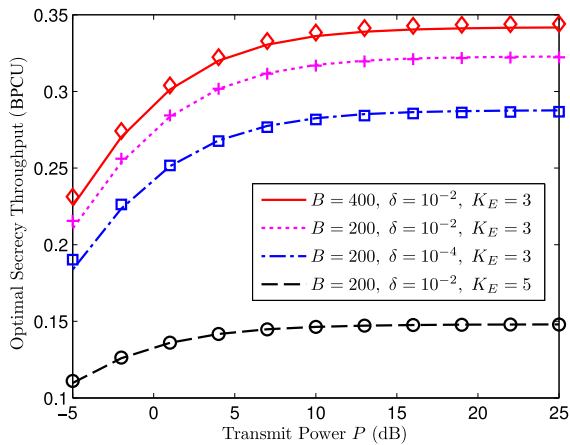}
	}
	\hfil
	\subfloat[]{\label{fig_opt_N}
		\includegraphics[width=3in]{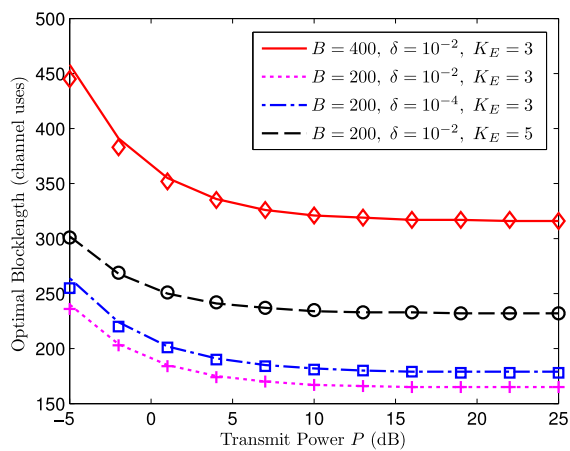}
	}
	\caption{(a) The optimal secrecy throughput and (b) the corresponding optimal blocklength versus the transmit power $P$ at the AP under different system parameters with $ \rho_A/P=10 $ dB and $ \rho_E/P=3 $ dB, where the lines are obtained through the proposed analytical approximation method in Theorem \ref{theorem_N} and Proposition \ref{prop_T2} while the markers are attained from the simulations and one-dimensional search.
	}
	\label{fig_opt}
\end{figure}

Fig. \ref{fig_opt} plots the optimal secrecy throughput and the corresponding optimal blocklength maximizing the throughput obtained from both the analytical and numerical methods.
It can be seen from  Fig. \ref{fig_opt} that the analytical results coincide well with the simulations.
Moreover, both the optimal secrecy throughput and the optimal blocklength remain fixed as the transmit power $ P $ at the AP becomes larger in the high-SNR regime. This is because the secrecy throughput is no longer a function of $ P $ in the high-SNR regime according to the analysis in Section \ref{subsec_high_SNR}.
As the system parameters change, their impacts on the optimum align well with our theoretical findings in Corollaries \ref{corollary_N} and \ref{corollary_T} from Fig. \ref{fig_opt}.

\begin{figure}[t]
	\centering
	\includegraphics[width=3in]{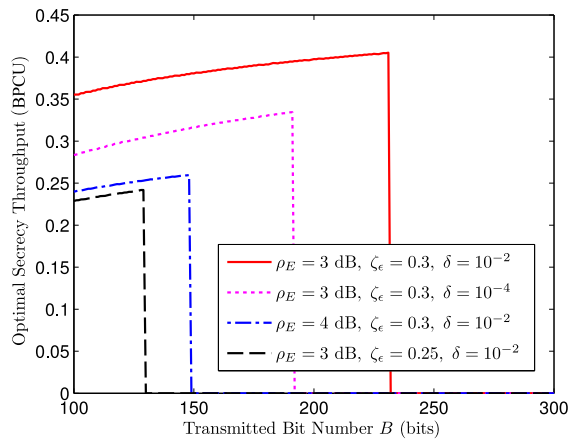}
	\caption{The optimal average secrecy throughput versus the transmitted bit number per block $ B $ under the reliability and latency constraints, where $ K_E=1 $ and $ \zeta_N=400 $ channel uses.}
	\label{fig_opt_T_cons}
\end{figure}

Fig. \ref{fig_opt_T_cons} depicts the optimal average secrecy throughput obtained from Corollary \ref{theorem_N1} for each fixed transmitted bit number per block $ B $ under the reliability and latency constraints, and the throughput is set to zero once the problem is infeasible.
From Fig. \ref{fig_opt_T_cons} one can observe that there exists one critical point and when $ B $ exceeds this point the optimal secrecy throughput returns to zero.
This is because, in order to maintain a fixed decoding error probability $ \zeta_\epsilon $, the blocklength $ N $ should be increased when a larger $ B $ is demanded.
However, $ N $ cannot be arbitrarily large due to the imposed latency constraint. Therefore, once $ B $ is larger than its critical point it will be infeasible to find a blocklength satisfying both the reliability and latency constraints.
%This is because the term $ \bar{\epsilon}^{-1} (\zeta_\epsilon) $ in problem \eqref{p1} increases with $ B $, and once $ B $ is such large that $ \left\lceil\bar{\epsilon}^{-1} (\zeta_\epsilon)\right\rceil > \left\lfloor\zeta_N\right\rfloor $ holds it will be infeasible to find a blocklength satisfying both the reliability and latency constraints.
Furthermore, it can be observed from Fig. \ref{fig_opt_T_cons} that both the optimal secrecy throughput and the critical point become smaller, when the secrecy or reliability constraints become more stringent or $ \rho_E $ increases. The reason behind this is that the blocklength is expected to be larger so as to meet the reliability constraint at this time, which considerably reduces the secrecy information rate.

\subsection{Multi-Antenna Case}\label{subsec_sim_multi}
\begin{figure}[t]
	\centering
	\includegraphics[width=3in]{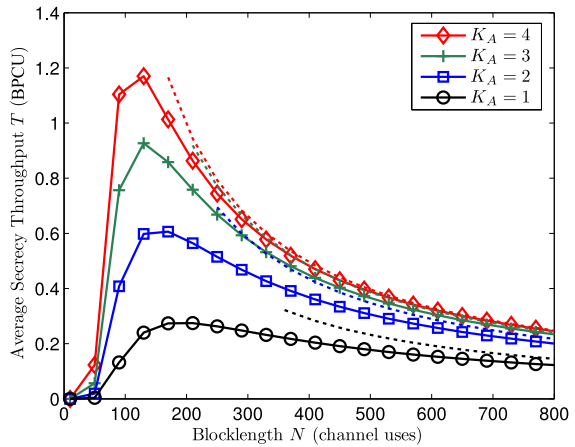}
	\caption{The average secrecy throughput $ T $ versus blocklength $ N $ with different antenna numbers at the AP $ K_A $, where the dashed lines denote the secrecy throughput for the asymptotic case with infinite blocklength and $ \eta=1 $.} %asymptotic infinite blocklength case
	\label{fig_T_Ka}
\end{figure}

In Fig. \ref{fig_T_Ka}, the impact of the antenna number of the AP $ K_A $ on the average secrecy throughput under the MRT beamforming scheme ($ \eta=1 $) is presented.
The dashed lines in Fig. \ref{fig_T_Ka} denote the secrecy throughput for the asymptotic case with infinite blocklength, which are evaluated according to $ T^{N\to \infty}=\frac{B}{N}\P\{\gamma_A>\gamma_E\} $ as in Section \ref{subsec_large_N}. According to Fig. \ref{fig_T_Ka}, when $ N>600 $ the secrecy throughput with finite blocklength coding coincides well with its infinite-blocklength counterpart especially for $ K_A\geq 3 $. In light of this observation, we know that adding more antennas at the AP can compensate for the performance loss incurred by short-packet communications in a way. % to some extent.
Moreover, it can be observed from Fig. \ref{fig_T_Ka} that increasing $ K_A $ can significantly improve the secrecy throughput. In particular, the maximum secrecy throughput gets approximately doubled when the antenna number of the AP increases from one to two, which shows the significant performance improvement provided by the extra spatial degrees of freedom.
Fig. \ref{fig_T_Ka} also shows that the optimal blocklength for secrecy throughput maximization decreases as $ K_A $ increases.
The reason behind this is that reducing transmission latency so as to maximize the secrecy rate is more favorable than mitigating the decoding error, when the AP can beam at the actuator using MRT and reduce the information leakage to the eavesdropper with more equipped antennas.

\begin{figure}[t]
	\centering
	\includegraphics[width=3in]{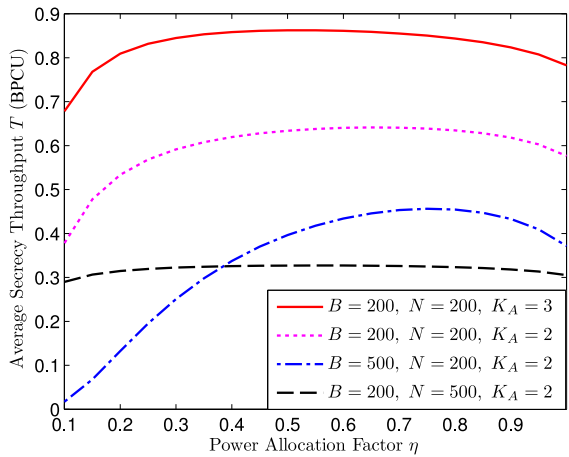}
	\caption{The average secrecy throughput $ T $ versus the power allocation factor $ \eta $ in the multi-antenna case.}
	\label{fig_T_eta}
\end{figure}

Fig. \ref{fig_T_eta} shows the impact of the power allocation on the secrecy throughput in the multi-antenna case with the AN-aided transmission scheme.
It can be seen that the power allocation factor $ \eta $ indeed plays an important role in maximizing the secrecy throughput and there exits an optimal $ \eta^* $ to maximize the throughput.
From Fig. \ref{fig_T_eta}, $ \eta^* $ increases as the total transmitted bit number per block $ B $ becomes larger, while it decreases as the antenna number of the AP $ K_A $ and the blocklength $ N $ increase.
The reason behind this is that the decoding error probability is a decreasing function of $ K_A $ and $ N $ while it increases as $ B $ becomes larger.
When the decoding error probability is large, it is favorable to allocate more power to the information-bearing signal rather than the AN to mitigate the decoding error.
Therefore, $ \eta^* $ increases when the decoding performance deteriorates.
On the contrary, when a smaller decoding error probability is met, $ \eta^* $ is expected to decrease so as to inject more AN signal for secrecy enhancement.

\section{Conclusion} \label{sec_conclusion}
In this paper, we have presented a comprehensive study on the average secrecy throughput of short-packet communications for secure IoT applications with an external multi-antenna eavesdropper.
In particular, an analytical framework has been proposed to approximate the average secrecy throughput with finite blocklength.
Specifically, we have first investigated the case with a single-antenna AP, where the secrecy throughput has been approximated in the closed form.
Based on the closed-form expression, we have analytically obtained the optimal blocklength in terms of secrecy throughput maximization, and the impacts of the system parameters as well as the reliability and latency constraints on the optimum have been further analyzed.
Moreover, the proposed theoretical approximation method has also been extended to the scenario where the AP has multiple antennas and adopts an AN-aided transmission scheme.
%Moreover, the proposed theoretical approximation method has been extended to the scenario where the AP has multiple antennas, and a closed-form expression for the secrecy throughput has been obtained under an AN-aided transmission scheme.
Numerical results verify the accuracy of the proposed approximation method and show that equipping multiple antennas at the transmitter can effectively improve the secrecy throughput and reduce the optimal blocklength in short-packet communications.
The performance of secure short-packet communications for mission-critical or uRLLC IoT applications has been comprehensively evaluated in our paper.
One promising future direction is to study the performance of secure mMTC IoT applications with large numbers of users. Another one is to introduce relaying nodes and investigate the reliability-latency tradeoff in secure cooperative IoT networks with short-packet transmission.

\begin{appendices}

\section{Proof of Proposition \ref{prop_T}}\label{app_prop_T_single}
Substituting \eqref{app_Psi2} into \eqref{T1} yields
\begin{align}\label{T2}
T\approx \frac{B}{N \rho_E^{K_E}\Gamma(K_E)} \int_{0}^{\infty}
y^{K_E-1} e^{-\left(\frac{x_0(y)}{\rho_A}+\frac{y}{\rho_E}\right)}
dy.
%T\approx \frac{B}{N} \int_{0}^{\infty}
%e^{-\frac{x_0(y)}{\rho_A}} \frac{1}{\rho_E}e^{-\frac{y}{\rho_E}}dy,
\end{align}
However, due to the complicated form of $ x_0(y) $ in \eqref{x0}, it is still hard to obtain a closed-form expression for \eqref{T2}. By noticing that $ V_E $ approaches $ 1 $ when $ \gamma_E $ becomes sufficiently large, we propose to further approximate the secrecy throughput in \eqref{T2} as follows:
\begin{align}\label{T_prop1}
T\approx& \frac{B}{N \rho_E^{K_E}\Gamma(K_E)} \Bigg( \underbrace{\int_{0}^{M_1}
	y^{K_E-1} e^{-\left(\frac{x_0(y)}{\rho_A}+\frac{y}{\rho_E}\right)} dy}_{\Omega_1} \notag\\
&\quad +
\underbrace{\int_{M_1}^{\infty} y^{K_E-1} e^{-\left(\frac{\varpi_1(1+y)-1}{\rho_A}+\frac{y}{\rho_E}\right)} dy}_{\Omega_2}
\Bigg).
\end{align}
By leveraging Gaussian-Chebyshev quadrature \cite{Hildebrand1987}, integral $ \Omega_1 $ can be approximated as follows:
\begin{align}\label{Omega1}
\Omega_1 \approx
\frac{M_1}{2} \sum_{n=1}^{M_2}\left(\varrho_{M_2} f\left(\frac{M_1}{2}(t_n+1)\right) \sqrt{1-t_n^2} \right).
\end{align}
According to \cite[eq. (3.351.2)]{Gradshteyn2007}, we have
\begin{align}\label{Omega2}
\Omega_2 = \frac{1}{\varpi_2^{K_E}} e^{-\frac{\varpi_1-1}{\rho_A}}\Gamma(K_E,\varpi_2 M_1).
\end{align}
The proof is complete by substituting \eqref{Omega1} and \eqref{Omega2} into \eqref{T_prop1}.

\section{Proof of Theorem \ref{theorem_N}}\label{app_theorem_N}
To find the optimal blocklength maximizing the secrecy throughput, we first relax the integer $ N $ as a positive real number.
Then the partial derivative of the secrecy throughput $ T $ in \eqref{app_T2} w.r.t. $ N $ can be found as follows:
\begin{align}\label{dT}
\frac{\partial T}{\partial N}=\frac{B}{N^2 \left(\rho_E\varpi_2\right)^{K_E}} e^{-\frac{\varpi_1-1}{\rho_A}} \cdot \Xi(N).
\end{align}
Since the factor $ \frac{B}{N^2 \left(\rho_E\varpi_2\right)^{K_E}} e^{-\frac{\varpi_1-1}{\rho_A}}>0 $ in \eqref{dT} for $ N\in (0,\infty) $, the sign of $ \frac{\partial T}{\partial N} $ only depends on that of $ \Xi(N) $.

As for function $ \Xi(N) $, we can easily check that it is a decreasing function of $ N\in (0,\infty) $ by noting that $ \frac{\varpi_1}{\varpi_2} $ is an increasing function of $ \varpi_1 $. Moreover, we have $ \lim\limits_{N \to 0^+ }\Xi(N)>0  $ while $ \lim\limits_{N\to \infty}\Xi(N)<0  $.
Therefore, we conclude that the secrecy throughput $ T $ first increases and then decreases with $ N\in (0,\infty)  $, which is thereby a quasi-concave function \cite[Sec. 3.4.2]{Boyd2004} of the continuous $ N $. The optimal $ N $ (the relaxed continuous value) maximizing $ T $ can be found as the unique zero-crossing point of $ \Xi(N) $.
The proof is complete.

\section{Proof of Corollary \ref{corollary_N}}\label{app_corollary_N}
According to the derivative rule for implicit functions with $ \Xi(N)=0 $, the impact of any parameter $ \chi $ on $ N^* $ can be analyzed through
\begin{align}
\frac{d N}{d \chi}= - \frac{\partial \Xi/\partial \chi}{\partial \Xi/\partial N}.
\end{align}
From Theorem \ref{theorem_N}, we already have $ \frac{\partial \Xi}{\partial N}<0 $. Therefore, the sign of $ \frac{d N}{d \chi} $ only depends on that of $ \frac{\partial \Xi}{\partial \chi} $.
The behavior $ N^* $ when the system parameters change is then discussed as follows:

\emph{1) $ N^* $ w.r.t. $ \delta $:} When $ \delta<0.5 $, we know that $ Q^{-1}(\delta)>0 $ is a decreasing function of $ \delta $. It is easy to show that $  \Xi(N)  $ is a decreasing function of $ \delta<0.5 $, i.e., $ \frac{\partial \Xi}{\partial \delta}<0 $, which thereby leads to $ \frac{d N}{d \delta}<0 $ for $ \delta<0.5 $.

\emph{2) $ N^* $ w.r.t. $ P $ and $ \rho_A $:} Since $ \Xi(\rho_A) \propto \frac{1}{\rho_A} \left(1+\frac{K_E}{\varpi_2}\right)=\frac{1}{\rho_A} +\frac{K_E}{\varpi_1+\rho_A/\rho_E} $, we have $ \frac{\partial \Xi}{\partial \chi}<0 $ for $ \chi\in\{P,\rho_A\} $, which thereby leads to $ \frac{d N}{d \chi}<0 $ for $ \chi\in\{P,\rho_A\} $.

\emph{3) $ N^* $ w.r.t. $ \rho_E $, $ K_E $, and $ B $:} From \eqref{N}, it can be easily seen that $  \Xi(N)  $ is an increasing function of $ \rho_E $, $ K_E $, and $ B $, i.e., $ \frac{\partial \Xi}{\partial \chi}>0 $ for $ \chi\in\{\rho_E,K_E,B\} $, which thereby leads to $ \frac{d N}{d \chi}>0 $ for $ \chi\in\{\rho_E,K_E,B\} $.

\section{Proof of Corollary \ref{corollary_T}}\label{app_corollary_T}
According to the chain rule of derivative, the impact of any parameter $ \chi $ on $ T^*\triangleq T(N^*(\chi),\chi) $ can be analyzed through
\begin{align}\label{chain}
\frac{d T^*}{d \chi}= \frac{\partial T}{\partial N}\Bigg|_{N=N^*} \times \frac{d N^*}{d \chi} + \frac{\partial T}{\partial \chi}\Bigg|_{N=N^*}=\frac{\partial T}{\partial \chi} \Bigg|_{N=N^*},
\end{align}
where the second equality follows from the fact that $ \frac{\partial T}{\partial N}\Big|_{N=N^*}=0 $ according to Theorem \ref{theorem_N}.
Therefore, it suffices to determine the sign of $ \frac{\partial T(N)}{\partial \chi} \Big|_{N=N^*} $, and the impact of the system parameters on $ T(N^*) $ is then discussed as follows:

\emph{1) $ T(N^*) $ w.r.t. $ \delta $:} Since $ \varpi_1 $ is a decreasing function of $ \delta $, it is easy to show that $ T $ increases with $ \delta $, i.e., $ \frac{\partial T}{\partial \delta}>0 $ for any $ N>0 $, which thereby leads to $ \frac{d T^*}{d \delta}>0 $.

\emph{2) $ T(N^*) $ w.r.t. $ P $ and $ \rho_A $:} When $ \delta<0.5 $, it is easy to check from \eqref{app_T2} that $ T $ is an increasing function of $ \chi\in\{P,\rho_A\} $ since the condition $ \varpi_1-1>0 $ holds for this case. Therefore, we have $ \frac{d T^*}{d \chi}>0 $ for $ \chi\in\{P,\rho_A\} $ when $ \delta<0.5 $.

\emph{3) $ T(N^*) $ w.r.t. $ B $:} The partial derivative of $ T $ w.r.t. $ B $ is first calculated as follows:
\begin{align} \label{T_B}
\frac{\partial T}{\partial B}=\frac{1}{N \left(\rho_E\varpi_2\right)^{K_E}} e^{-\frac{\varpi_1-1}{\rho_A}}
\left(1- \frac{\varpi_1}{\rho_A} \left(1+\frac{K_E}{\varpi_2}\right) \frac{B}{N} \ln 2  \right).
\end{align}
According to the fact that $ \Xi(N^*)=0 $ from Theorem \ref{theorem_N}, we further obtain
\begin{align}
&\frac{\partial T}{\partial B}\Bigg|_{N=N^*}  \notag\\
&=\frac{1}{N \left(\rho_E\varpi_2\right)^{K_E}} e^{-\frac{\varpi_1-1}{\rho_A}}
\frac{\varpi_1}{\rho_A} \left(1+\frac{K_E}{\varpi_2}\right) \frac{Q^{-1}(\delta)}{2\sqrt{N}}\Bigg|_{N=N^*}>0,
\end{align}
which finally yields $ \frac{d T^*}{d B}>0 $ from \eqref{chain}.

\emph{4) $ T(N^*) $ w.r.t. $ \rho_E $ and $ K_E $:} From \eqref{app_T2}, it can be easily seen that $  T  $ is a decreasing function of $ \rho_E $ and $ K_E $, i.e., $ \frac{\partial T}{\partial \chi}<0 $ for any $ N>0 $ and $ \chi\in\{\rho_E,K_E\} $, which thereby leads to $ \frac{d T^*}{d \chi}<0 $ for $ \chi\in\{\rho_E,K_E\} $.

\section{Proof of Theorem \ref{theorem_T_multi}}\label{app_theorem_T_multi}
By using partial integration, the secrecy throughput in \eqref{T_m} can be further calculated as follows:
\begin{align}\label{T_m1}
T=& -\frac{B}{N} \int_{0}^{\infty} F_{\gamma_E}(y) d\Psi(y) \notag\\
=&\frac{B}{N} \Bigg(  \underbrace{\int_{0}^{M_1} F_{\gamma_E}(y) f_{\gamma_A}\left(x_0(y)\right) x'_0(y) dy}_{\Phi_1} \notag\\
&\qquad +
\underbrace{\int_{M_1}^{\infty} F_{\gamma_E}(y) f_{\gamma_A}\left(x_0(y)\right) x'_0(y) dy}_{\Phi_2} \Bigg),
%\overset{(b)}{\approx}& \frac{B}{N}
\end{align}
where $ M_1 $ is a sufficiently large parameter as in the single-antenna case.

The integral $ \Phi_1 $ in \eqref{T_m1} can be approximated via Gaussian-Chebyshev quadrature as
\begin{align}\label{Phi_1}
\Phi_1 \approx \frac{M_1}{2} \sum_{n=1}^{M_2}\left(\varrho_{M_2} g\left(\frac{M_1}{2}(t_n+1)\right) \sqrt{1-t_n^2} \right).
\end{align}
Since $ M_1 $ is sufficiently large, we approximate $ x_0(y) $ and $ x'_0(y) $ in integral $ \Phi_2 $ of \eqref{T_m1} as $ x_0(y)\approx \varpi_1(1+y)-1 $ and $ x'_0(y) \approx\varpi_1 $, respectively.
Then, the integral $ \Phi_2 $ can be approximated as 
\begin{align}\label{Phi_2}
\Phi_2\approx& \int_{M_1}^{\infty} F_{\gamma_E}(y) f_{\gamma_A}\left(\varpi_1(1+y)-1\right) \varpi_1 dy \notag\\
=& \frac{1}{\Gamma(K_A)}\Gamma\left(K_A,\frac{\varpi_1(M_1+1)-1}{\eta\rho_A}\right)-
\frac{\varpi_1 e^{-\frac{\varpi_1-1}{\eta\rho_A}}}{(\eta\rho_A)^{K_A}\Gamma(K_A)}
\Phi_3,
\end{align}
where
\begin{align}\label{Phi_3}
\Phi_3\triangleq &
\int_{M_1}^{\infty}
(\varpi_1 y+\varpi_1-1)^{K_A-1}e^{- \varpi_3 y}
\sum_{n=1}^{K_E} \frac{A_n(y)}{(n-1)!}\left(\frac{y}{\eta\rho_E}\right)^{n-1}
dy \notag\\
=&
\sum_{p=0}^{K_A-1}\binom{K_A-1}{p}  \left(\frac{\varpi_1}{\varpi_1-1}\right)^p
\sum_{n=1}^{K_E} \frac{(\varpi_1-1)^{K_A-1}}{(n-1)!(\eta\rho_E)^{n-1}} \notag\\
&\qquad \times
\underbrace{\int_{M_1}^{\infty} A_n(y) y^{n+p-1} e^{- \varpi_3 y} dy}_{\Theta_n}
\end{align}
with $ \varpi_3 \triangleq  \frac{\varpi_1 \rho_E+\rho_A}{\eta\rho_A\rho_E}$.

%By using the binomial theorem,
According to different choices of $\eta$, $K_A$, and $K_E$, the calculation of integral $ \Theta_n $ in \eqref{Phi_3} can be further discussed as follows:

\emph{1) The case of $ \eta=1 $:} In this case, the AP allocates all the transmitting power to the information-bearing signal and there is no injected AN. Since $ \tau=0 $ under this situation, we always have $ A_n(x)=1 $ according to \eqref{An}. By leveraging \cite[eq. (3.351.2)]{Gradshteyn2007}, integral $ \Theta_n $ is calculated as follows:
\begin{align}
\Theta_n =\Theta_{1,n}\triangleq
\frac{1}{\varpi_3^{n+p}}\Gamma(n+p,\varpi_3 M_1),\quad \text{for}~1\leq n \leq K_E.
\end{align}

\emph{2) The case of $ \eta\neq 1 $ and $ K_E<K_A $:} When the antenna number of the AP is larger than that at the eavesdropper, we have $ A_n(x)=\frac{\sum_{m=0}^{K_E-n}\binom{K_A-1}{m}(\tau x)^m}{(1+\tau x)^{K_A-1}} $ for $ 1\leq n \leq K_E $ from \eqref{An}.
The integral $ \Theta_n $ for $ 1\leq n \leq K_E $ now becomes
\begin{align}
\Theta_n &=\Theta_{2,n}\notag\\
&\triangleq
\sum_{m=0}^{K_E-n}\binom{K_A-1}{m} \tau^m
\int_{M_1}^{\infty} \frac{y^{p+m+n-1}}{(1+\tau y)^{K_A-1}} e^{- \varpi_3 y} dy \notag\\
&\overset{(a)}{=} \sum_{m=0}^{K_E-n}\binom{K_A-1}{m} \frac{\tau^m e^{-\varpi_3 M_1}}{(1+\tau M_1)^{K_A-1}}\notag\\
& \qquad \times
\sum_{q=0}^{p+m+n-1}\binom{p+m+n-1}{q} M_1^{p+m+n-1-q} \Phi_4,
\end{align}
where step $ (a) $ follows from the variable change $ z=y-M_1 $ and the binomial theorem, and
\begin{align}\label{Phi_4}
&\Phi_4\notag\\
&\triangleq 
\int_{0}^{\infty} \frac{z^q}{(1+\frac{\tau}{1+\tau M_1} z)^{K_A-1}} e^{- \varpi_3 y} dz \notag\\
&\overset{(b)}{=}
\left(\frac{1+\tau M_1}{\tau}\right)^{q+1}
\int_{0}^{\infty} \frac{t^q}{(1+t)^{K_A-1}} e^{- \frac{1+\tau M_1}{\tau} \varpi_3 t} dt  \notag\\
&\overset{(c)}{=}
\left(\frac{1+\tau M_1}{\tau}\right)^{q+1}
\Gamma(q+1) U\left(q+1,q-K_A+3,\frac{1+\tau M_1}{\tau} \varpi_3\right)
\end{align}
where steps $ (b) $ and $ (c) $ follow from the variable change $ t=\frac{\tau}{1+\tau M_1} z $ and \cite[eq. (9.211.4)]{Gradshteyn2007}, respectively.
%Note that $ U(\cdot,\cdot,\cdot) $ in \eqref{Phi_4} is the Tricomi's (confluent hypergeometric) function, which can be easily evaluated by computational software.

\emph{3) The case of $ \eta\neq 1 $ and $ K_E\geq K_A $:} When the eavesdropper has more antennas than the AP, according to \eqref{An} the integral $ \Theta_n $ now takes the form of
\begin{align}
\Theta_n =
\begin{cases}
\Theta_{1,n},&1\leq n\leq K_E-K_A+1,\\
\Theta_{2,n},&K_E-K_A+2\leq n\leq K_E.
\end{cases}
\end{align}
The proof is complete by substituting \eqref{Phi_1}--\eqref{Phi_3} into \eqref{T_m1}.
\end{appendices}

\bibliographystyle{IEEEtran}

\end{document}